\documentclass[conference]{IEEEtran}
\IEEEoverridecommandlockouts
\usepackage{cite}
\usepackage{amsmath,amssymb,amsfonts}
\usepackage{algorithmic}
\usepackage{graphicx}
\usepackage{textcomp}
\usepackage{xcolor}
\usepackage[hidelinks]{hyperref}
\def\BibTeX{{\rm B\kern-.05em{\sc i\kern-.025em b}\kern-.08em
    T\kern-.1667em\lower.7ex\hbox{E}\kern-.125emX}}

\usepackage{float}
\begin{document}

\title{Attack-Specialized Deep Learning with Ensemble Fusion for Network Anomaly Detection}

\author{\IEEEauthorblockN{Nisith Dissanayake}
\IEEEauthorblockA{\textit{Department of Computer Science and Engineering} \\
\textit{University of Moratuwa}\\
Colombo, Sri Lanka\\
nisith.21@cse.mrt.ac.lk}
\and
\IEEEauthorblockN{Dr. Uthayasanker Thayasivam}
\IEEEauthorblockA{\textit{Department of Computer Science and Engineering} \\
\textit{University of Moratuwa}\\
Colombo, Sri Lanka\\
rtuthaya@cse.mrt.ac.lk}
}

\maketitle

\begin{abstract}
The growing scale and sophistication of cyberattacks pose critical challenges to network security, particularly in detecting diverse intrusion types within imbalanced datasets. Traditional intrusion detection systems (IDS) often struggle to maintain high accuracy across both frequent and rare attacks, leading to increased false negatives for minority classes. To address this, we propose a hybrid anomaly detection framework that integrates specialized deep learning models with an ensemble meta-classifier. Each model is trained to detect a specific attack category, enabling tailored learning of class-specific patterns, while their collective outputs are fused by a Random Forest meta-classifier to improve overall decision reliability. The framework is evaluated on the NSL-KDD benchmark, demonstrating superior performance in handling class imbalance compared to conventional monolithic models. Results show significant improvements in precision, recall, and F1-score across all attack categories, including rare classes such as User to Root (U2R). The proposed system achieves near-perfect detection rates with minimal false alarms, highlighting its robustness and generalizability. This work advances the design of intrusion detection systems by combining specialization with ensemble learning, providing an effective and scalable solution for safeguarding modern networks.
\end{abstract}


The proposed framework was implemented in Python and trained on the NSL-KDD dataset.\footnote{Source code and dataset: \url{https://github.com/aaivu/In21-S7-CS4681-AML-Research-Projects/tree/main/projects/210144G-Cybersecurity-AI_Threat-Detection}}

\section{Introduction}
The increasing intricacy of the digital landscape has made cybersecurity a critical global concern. As cyber threats become more complex, traditional, signature-based security solutions are often inadequate, as they struggle to detect sophisticated, shape-shifting attacks \cite{ugochukwu_ikechukwu_okoli_machine_2024}. This has created an urgent need for more advanced and flexible security strategies. The field of intrusion detection systems (IDS) is at the forefront of this effort, with a growing focus on leveraging modern machine learning (ML) and deep learning (DL) techniques to analyze vast amounts of network data and identify malicious activities \cite{ugochukwu_ikechukwu_okoli_machine_2024}.

Unlike conventional methods that rely on predefined rules, ML algorithms can identify subtle patterns and anomalies within large datasets, offering a more effective approach to threat detection \cite{ugochukwu_ikechukwu_okoli_machine_2024}. This capability is particularly vital for anomaly-based IDS, which are designed to detect novel attacks by flagging any behavior that deviates from a learned "normal" baseline \cite{shana_anomaly-based_2025, noauthor_pdf_nodate-1}. However, training these models can be resource-intensive, and their performance is often dependent on the quality and balance of the training data \cite{mirsky_kitsune_2018}.

The NSL-KDD dataset has emerged as a widely used benchmark for evaluating network intrusion detection systems, addressing some of the statistical shortcomings of its predecessor, the KDD Cup 99 dataset \cite{tavallaee_detailed_2009, li_comparison_2018}. The dataset includes a variety of attacks, such as \textbf{Denial of Service (DoS)}, \textbf{Probing (Probe)}, \textbf{User-to-Root (U2R)}, and \textbf{Remote-to-Local (R2L)}, making it a valuable tool for testing the effectiveness of new IDS models \cite{khanam_towards_2022}. Despite its widespread use, the inherent challenges of imbalanced data and the need for high-performance, real-time detection remain \cite{mulyanto_effectiveness_2020, khanam_towards_2022}.

The core challenge in building an effective NIDS is the diverse and heterogeneous nature of network attacks. Attack types, such as Denial of Service (DoS), Probe, Remote-to-Local (R2L), and User-to-Root (U2R), possess distinct characteristics. For example, DoS attacks are high-volume, flooding-based events, while U2R attacks are low-volume, stealthy infiltrations that are extremely difficult to detect. Compounding this issue is the severe class imbalance, where attacks constitute a small fraction of the total network traffic, particularly for rare attack types like R2L and U2R. A single, monolithic model designed to detect all these attack types simultaneously often performs sub-optimally, as it struggles to learn both the high-volume patterns and the subtle, low-volume anomalies.

This paper presents a novel deep ensemble intrusion detection system designed to address these challenges. It proposes a specialized architecture that goes beyond traditional single-model approaches by employing individual binary detectors for each major attack family present in the NSL-KDD dataset. By combining the strengths of various ML techniques, including deep learning models like \textbf{Convolutional Neural Networks (CNNs)} and \textbf{Long Short-Term Memory (LSTM)} networks, \textbf{Attention Mechanism} and traditional ML algorithms like \textbf{Random Forest (RF)}, the system is engineered to accurately distinguish between normal and malicious traffic. The system's ability to maintain high performance on both the KDDTest+ and KDDTest-21 subsets demonstrates its generalizability and robustness against real-world distribution shifts.

\section{Related Work}

Machine learning models fundamentally reshape cybersecurity strategies by enabling adaptive, proactive, and predictive defense postures \cite{sailpoint_how_2025}. By continuously processing and analyzing immense volumes of network traffic data, these models can identify subtle anomalies and complex patterns in real time, thereby allowing for the detection of novel threats before they can inflict significant damage \cite{sailpoint_how_2025, comparitech_machine_2025}. This approach moves beyond the limitations of a closed-world system, which can only identify what it has been explicitly trained to recognize, toward an open-world model capable of generalizing from learned patterns to new data \cite{mirsky_kitsune_2018}. Furthermore, the application of ML extends beyond mere detection, driving the automation of threat response. Machine learning-based systems can automatically initiate predefined mitigation actions such as isolating affected systems or blocking malicious IP addresses within seconds of a threat being identified, minimizing potential harm and expediting incident response \cite{sailpoint_how_2025}. This fundamental change in methodology, from a reactive, rule-based defense to a predictive, behavior based one, represents a critical advancement in the field of network security.

Intrusion Detection Systems (IDS) are essential components of a layered defense architecture, functioning as a device or software application that continuously monitors a network or system for signs of malicious activity or policy violations \cite{wikipedia_intrusion_2025}. Their primary purpose is to identify potential threats and provide an early warning to system administrators, capturing and logging critical information that can be used for later investigation of a data breach \cite{dhs_intrusion_2025}. The field of IDS has historically been dominated by two principal methodologies: signature-based detection and anomaly-based detection \cite{alashjaee_hybrid_2023, aljanabi_effective_2023}.

\textbf{Signature-based IDS}, also known as misuse detection, operates on a principle similar to traditional antivirus software. It maintains a database of known intrusion signatures, and any network traffic that matches a signature is flagged as an attack \cite{alashjaee_hybrid_2023}. This method is highly accurate and yields a low rate of false alarms for attacks whose signatures are contained within the database \cite{alashjaee_hybrid_2023}. However, its dependency on a finite database renders it ineffective against new and evolving attacks for which no signatures exist, leading to a high rate of false negatives for novel threats \cite{alashjaee_hybrid_2023}. In contrast, \textbf{Anomaly-Based IDS} establishes a baseline of normal network behavior by profiling the activities of users and systems \cite{alashjaee_hybrid_2023}. Any deviation from this established baseline is identified as anomalous and, therefore, potentially malicious. This approach is powerful because it can detect previously unknown intrusions, making it effective against zero-day attacks \cite{alashjaee_hybrid_2023}. The main drawback of anomaly-based systems is their potential for a higher rate of false alarms, as they may misclassify legitimate but unusual behavior as a threat \cite{alashjaee_hybrid_2023}. The inherent trade-off between the precision of signature-based systems and the generality of anomaly-based systems has created a demand for more sophisticated, hybrid solutions that can mitigate the weaknesses of each approach. This need has been the catalyst for the development of advanced deep learning models that can learn to identify both known attack patterns and general behavioral anomalies.

The limitations of traditional IDS and the complexity of modern network data have driven the adoption of deep learning, with hybrid architectures proving to be exceptionally effective. The combination of convolutional neural networks (CNNs) and long short-term memory (LSTM) networks is particularly prominent due to its unique ability to process the bi-faceted nature of network traffic.

The success of the hybrid CNN-LSTM architecture for intrusion detection stems from its design, which mirrors the intrinsic structure of network traffic data. Network traffic can be conceptualized as having both static and dynamic characteristics. The CNN component is uniquely suited to detect spatial patterns of network traffic data by acting as a feature extractor \cite{bamber_hybrid_2025}. It analyzes the relationships and hierarchies among features in a single data record, identifying local patterns indicative of malicious activity.

Following this initial feature extraction, the LSTM component takes over. Network traffic is not a static collection of records but a continuous stream of sequential data where dependencies exist across time. The LSTM, a specialized type of recurrent neural network (RNN), is designed to model sequential dependencies across time and analyze the dynamic aspects of this data stream \cite{bamber_hybrid_2025}. By processing the sequence of features extracted by the CNN, the LSTM can identify temporal patterns that may indicate a sophisticated attack unfolding over a longer duration. This synergy provides a comprehensive and robust framework for distinguishing between benign and malicious network behavior, leading to higher accuracy and a reduced false positive rate than either model could achieve in isolation \cite{bamber_hybrid_2025, alashjaee_hybrid_2023}. This hybrid approach has demonstrated superior performance on diverse datasets like UNSW-NB15 and X-IIoTID, achieving high accuracies such as 93.21\% and 99.84\% respectively for binary classification \cite{alashjaee_hybrid_2023}.

While the CNN-LSTM architecture provides a strong foundation for intrusion detection, its performance can be further refined by incorporating an attention mechanism \cite{alashjaee_deep_2025}. This enhancement addresses a key challenge in network data analysis: the presence of vast amounts of heterogeneous and potentially irrelevant data that can obscure the subtle patterns of an attack.

An attention mechanism is a powerful technique that allows a model to highlight the utmost informative input features \cite{alashjaee_deep_2025}. In the context of a hybrid IDS, this mechanism enables the model to dynamically assign a weight to each feature and time step based on its significance to the detection task \cite{alashjaee_deep_2025}. This is a fundamental improvement over traditional architectures where all features contribute equally, a design that can dilute the relevance of a few critical attack indicators that are often buried within a large volume of non-malicious data \cite{alashjaee_deep_2025}. The attention layer acts as a computational focusing lens, enabling the model to pay more attention to the elements of the input that are most indicative of a threat \cite{nguyen_attention_2025}. By directing the model's focus to important attack features while effectively ignoring unnecessary data, the attention mechanism significantly enhances the system's ability to identify malicious activities, particularly in high-dimensional or noisy datasets \cite{alashjaee_deep_2025}. Empirical studies confirm the value of this addition, showing that Attention-CNN-LSTM models can achieve 94.8–97.5\% accuracy on datasets like NSL-KDD and Bot-IoT \cite{alashjaee_deep_2025}. An ablation study has further confirmed that the attention layer alone contributed to a 3-4\% improvement in F1-score and MCC \cite{alashjaee_deep_2025}, demonstrating its substantial impact on a model's discriminative power.

\section{Methodology: An Attack-Specialized Hybrid Framework}
\subsection{Data Acquisition and Preprocessing Pipeline}
The foundational step of this research involved establishing a robust and standardized preprocessing pipeline to ensure data consistency across all specialized models. Data was acquired from the NSL-KDD dataset \cite{tavallaee_detailed_2009}, which is provided in distinct training and test CSV files. The pipeline, applied uniformly to both datasets, addresses the unique challenges posed by the NSL-KDD's mixed-type feature set.

\begin{table}[h]
\centering
\caption{Class distribution in the NSL-KDD}
\begin{tabular}{lrr}
\hline
\textbf{Class} & \textbf{Count} & \textbf{Percentage} \\
\hline
Normal & 67,342 & 53.46\% \\
DoS    & 45,927 & 36.46\% \\
Probe  & 11,656 & 9.25\%  \\
R2L    &    995 & 0.79\%  \\
U2R    &     52 & 0.04\%  \\
\hline
\end{tabular}
\label{tab:class_distribution}
\end{table}

As shown in Table~\ref{tab:class_distribution}, the NSL-KDD dataset suffers from a severe class imbalance. The majority of the samples belong to the \textit{Normal} and \textit{DoS} classes (together accounting for nearly 90\%), while critical classes such as \textit{R2L} (0.79\%) and \textit{U2R} (0.04\%) are extremely underrepresented. This imbalance poses a significant challenge, as models trained without addressing it risk being biased toward majority classes and failing to detect minority attacks, which are often the most critical in intrusion detection contexts.

A crucial part of the preprocessing pipeline was the handling of categorical features, which included \texttt{protocol\_type}, \texttt{service}, \texttt{flag}, and the target variable, attack. An initial analysis was performed to identify the unique categories for each feature in both the training and test sets. This step was essential for preventing discrepancies in the feature space. The string-based categorical values were first converted into numerical representations using a LabelEncoder. Subsequently, OneHotEncoder (or DummyEncoder) was applied to transform these numerical labels into a one-hot encoded format. A critical consideration during this process was ensuring that the encoder was fit on a combined list of unique categories from both the training and test sets. This guarantees that the final one-hot encoded feature space is consistent, preventing issues where a category present in the test set but not the training set would be mishandled. The process successfully verified that the final categorical dataframes for both sets had a consistent column structure, which was a prerequisite for proceeding with model training.

Following the encoding of categorical data, a series of deliberate column-dropping operations were performed. The original string-based columns (\texttt{protocol\_type}, \texttt{service}, \texttt{flag}) were dropped after their one-hot encoded counterparts had been created and concatenated into a new dataframe. Furthermore, the \texttt{attack} column itself was dropped from the feature set (\texttt{X}) dataframes, leaving it solely as the target variable (\texttt{y}) for model training.

The final step in the preprocessing pipeline was feature scaling. The numerical features were standardized using StandardScaler from sklearn.preprocessing. The scaler was fitted exclusively on the training data and then used to transform both the training and test sets. This robust, standardized pipeline was applied consistently across all four specialized models, ensuring that any differences in performance could be attributed to the subsequent specialized techniques rather than variations in data preparation.

\subsection{Framework Architecture and Specialization}
The proposed framework is architected as a series of parallel, attack-specialized pipelines, each designed to detect a specific type of network intrusion. This design is a direct response to the heterogeneous nature of the NSL-KDD dataset, where different attack types exhibit dramatically different characteristics and frequencies. For instance, DoS attacks are abundant, while U2R and R2L attacks are extremely rare. A single model would be forced to learn conflicting patterns and would likely fail to perform adequately on the rarest classes. The specialized approach allows each pipeline to employ a unique combination of model architecture, feature selection, and class imbalance handling tailored to its specific task. The output of each specialized model is a probability score, which would then be aggregated by a final meta-classifier to produce the final classification decision.

\subsection{Specialized Model Training Pipelines}

\begin{table*}[htbp]
\centering
\caption{Summary of model architectures for each attack category.}
\label{tab:model_architecture}
\begin{tabular}{p{2cm} p{15cm}}
\hline
\textbf{Attack Type} & \textbf{Model Architecture Summary} \\
\hline
DoS & Conv1D(64,3) $\rightarrow$ Conv1D(64,3) $\rightarrow$ MaxPooling1D(2) $\rightarrow$ Conv1D(128,3)$\times$2 $\rightarrow$ MaxPooling1D(2) $\rightarrow$ BatchNorm $\rightarrow$ LSTM(100, dropout=0.1) $\rightarrow$ Dropout(0.5) $\rightarrow$ Dense(1, Sigmoid) \\
\hline
Probe & 
\textbf{CNN--BiLSTM--Attention}: Conv1D(64,3) $\rightarrow$ BatchNorm $\rightarrow$ Residual Block $\rightarrow$ MaxPooling1D(2) $\rightarrow$ Dropout(0.2) $\rightarrow$ Conv1D(128,3)$\times$2 $\rightarrow$ BatchNorm $\rightarrow$ MaxPooling1D(2) $\rightarrow$ Dropout(0.3) $\rightarrow$ Multi-Head Attention(4 heads, key\_dim=32) $\rightarrow$ BiLSTM(64,32) $\rightarrow$ Dense(128 $\rightarrow$ 64) $\rightarrow$ Dropouts $\rightarrow$ Dense(1, Sigmoid). \\ 
& \textbf{Ensemble models}: Random Forest (200 trees, max\_depth=10, class\_weight=\{0:1, 1:5\}), Gradient Boosting (100 estimators, depth=6, learning\_rate=0.1), Logistic Regression (max\_iter=1000, class\_weight=\{0:1, 1:4\}). \\ 
& \textbf{Ensemble layer}: Weighted sum of predicted probabilities: 0.4$\times$CNN--LSTM + 0.3$\times$RF + 0.2$\times$GB + 0.1$\times$LR. \\
\hline
R2L & Conv1D(64,3) $\rightarrow$ MaxPooling1D(2) $\rightarrow$ Conv1D(128,3) $\rightarrow$ MaxPooling1D(2) $\rightarrow$ LSTM(100, dropout=0.1) $\rightarrow$ Dense(1, Sigmoid). Loss: focal loss with cost-sensitive learning. Optimizer: Adam with adaptive learning rate. \\
\hline
U2R & Conv1D(32,3) $\rightarrow$ MaxPooling1D(2) $\rightarrow$ Dropout(0.2) $\rightarrow$ Conv1D(64,3) $\rightarrow$ MaxPooling1D(2) $\rightarrow$ Dropout(0.3) $\rightarrow$ LSTM(32, dropout=0.2, recurrent\_dropout=0.2) $\rightarrow$ Dense(16, ReLU) $\rightarrow$ Dropout(0.4) $\rightarrow$ Dense(num\_classes, Sigmoid). \\
\hline
\end{tabular}
\end{table*}
\subsubsection{DoS Attack Detection Model}
The DoS attack detection model was designed to handle high-volume denial-of-service attacks. The core of this pipeline is a deep learning architecture, specifically a CNN-LSTM model. The architecture consists of a Conv1D layer for feature extraction, followed by a MaxPooling1D layer. The output is then passed to an LSTM layer to capture temporal dependencies, and finally to a Dense output layer. Dropout and BatchNormalization layers are strategically placed to prevent overfitting and improve training stability.

To address the class imbalance, the training data was augmented using SMOTE (Synthetic Minority Over-sampling Technique). Additionally, a custom focal\_loss function was implemented with specific gamma and alpha parameters. This loss function, designed for highly imbalanced datasets, places a greater emphasis on correctly classifying the minority DoS attack samples. The training process was controlled using several callbacks, including EarlyStopping to prevent overfitting, LearningRateScheduler to dynamically adjust the learning rate, and ModelCheckpoint to save the best-performing model. The performance was evaluated using standard metrics such as accuracy, precision, recall, and F1-score.

\subsubsection{Probe Attack Detection Model}
The Probe attack detection model focuses on reconnaissance activities, which are often subtle and precede a more serious intrusion. This pipeline requires a model highly sensitive to nuanced patterns. The specialized CNN-LSTM architecture for this task was enhanced with MultiHeadAttention and Bidirectional layers. The MultiHeadAttention mechanism allows the model to simultaneously focus on different parts of the input sequence, capturing complex relationships that might be missed by a standard CNN-LSTM. The Bidirectional layer processes the sequence both forwards and backward, providing a richer understanding of the temporal context.

Class imbalance was handled through a combination of SMOTE for data augmentation and an "enhanced focal loss" with a "recall boost". The recall boost is a crucial adaptation, as it specifically penalizes false negatives more heavily, prioritizing the detection of the subtle Probe attacks. The model's performance was further optimized through a detailed analysis of the Precision-Recall curve, which is a more informative metric for imbalanced data than accuracy. This analysis led to the identification of an optimal classification threshold that balanced precision and recall, as evidenced by the F1-score. This strategic use of metrics and threshold optimization demonstrates a sophisticated understanding of the practical requirements for a security-oriented detection system, where a missed attack can have severe consequences.

The original CNN--LSTM model achieved a recall of 0.7390, a precision of 0.8387, and an F1-Score of 0.7857, with 632 missed attacks. To improve performance, an ensemble was employed using a combination of classical models and the CNN--LSTM, as detailed in Table~\ref{tab:model_architecture}. The ensemble model resulted in a significant improvement, achieving a recall of 0.9335 (+19.5\%), a precision of 0.8845, an F1-Score of 0.9084, and reducing missed attacks to 161 (an improvement of 471).

\subsubsection{R2L Attack Detection Model}
R2L attacks, which involve gaining unauthorized local access from a remote machine, are characterized by an extreme class imbalance. The approach for R2L detection was designed to directly confront this imbalance. A CNN-LSTM architecture was again employed, but with a specific focus on mitigating the data scarcity.

To handle the extreme imbalance, the methodology employed a two-pronged approach. First, SMOTE was used to oversample the minority R2L class, creating synthetic samples to balance the dataset. Second, the model was trained using cost-sensitive learning, where custom class weights were computed and applied. This technique directly penalizes the model more for misclassifying the rare R2L class, effectively forcing it to prioritize the correct detection of these critical events. This method is a direct and targeted response to the fundamental challenge posed by R2L attacks. The application of SMOTE, focal loss, and a CNN-LSTM architecture for R2L detection represents a sophisticated and necessary step beyond more traditional methods, demonstrating the project's novel contribution to solving a notoriously difficult problem.

\subsubsection{U2R Attack Detection Model}
The U2R attack, where a user escalates privileges to gain root access, is arguably the most challenging of the four categories due to its extreme scarcity, representing only 0.07\% of the total dataset. A single, monolithic model would almost certainly fail to detect this class. The U2R detection pipeline was therefore designed with an extensive set of techniques to address this.

Instead of a single oversampling method, the approach explored multiple variants: SMOTE, ADASYN, and BorderlineSMOTE. This multi-faceted strategy accounts for the different ways these algorithms generate synthetic data, allowing the system to find the most effective method for this specific data distribution. A custom "extreme focal loss" function was also developed, which includes a "penalty factor" for missing attacks, further reinforcing the model's focus on the rare U2R class. To combat overfitting on the tiny number of real samples, a "lightweight" and "ultra-lightweight" CNN-LSTM architecture was a deliberate design choice. This smaller model footprint is less prone to memorizing the sparse data and generalizes better. The framework also introduced a "Super-Ensemble" model, which combines the predictions from multiple classifiers to achieve a more robust and accurate final result.

\subsection{The Meta-Classifier}
The outputs of the four specialized models for DoS, Probe, R2L, and U2R are intended to be fed into a meta-classifier. The core concept is that the raw outputs from each model (i.e., the probability of a sample belonging to each class) are concatenated into a new, higher-level feature vector. A final, simpler classifier, such as a Random Forest is then trained on this new feature vector. This final step synthesizes the specialized knowledge from each of the four models into a single, cohesive decision-making unit. The meta-classifier is responsible for making the final classification decision (e.g., whether a sample is Normal, or an attack type), completing the narrative of the hybrid framework.

\begin{figure}[h]
    \centering
    \includegraphics[width=0.75\linewidth]{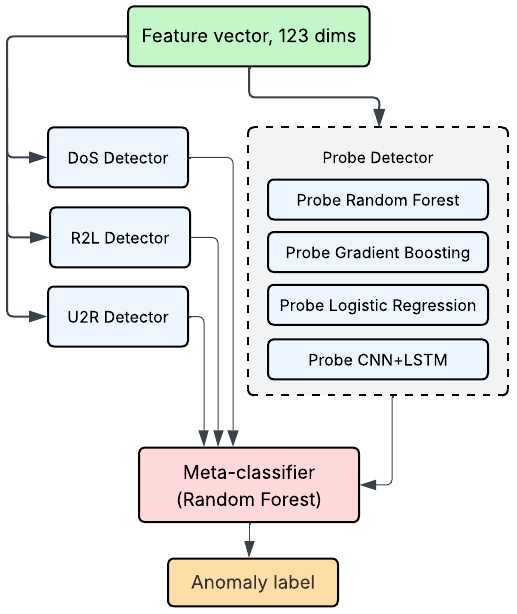}
    \caption{Attack Specific Multi Model Architecture}
    \label{fig:clahe}
\end{figure}

\section{Results}
\subsection{Evaluation Metrics}
The choice of evaluation metrics was critically important for this research, particularly given the severe class imbalance in the NSL-KDD dataset. It is well-established that \textbf{Accuracy} alone can be misleading in such cases, as a model can achieve high accuracy by simply classifying all samples as the majority class. Therefore metrics such as \textbf{Precision}, \textbf{Recall}, and the \textbf{F1-score} are important to get a good idea of the quality of results.

\textbf{Precision} measures the proportion of true positives among all positive predictions:
\[
P = \frac{\text{True Positives}}{\text{True Positives} + \text{False Positives}}
\]

\textbf{Recall} (also known as the true positive rate) measures the proportion of true positives among all actual positives:
\[
R = \frac{\text{True Positives}}{\text{True Positives} + \text{False Negatives}}
\]

\textbf{F1-score} provides a harmonic mean of precision and recall, offering a balanced measure of performance:
\[
\text{F1-score} = 2 \times \frac{P \times R}{P + R}
\]

In addition, the \textbf{Confusion Matrix} was employed to visualize the number of true positives, false positives, true negatives, and false negatives, providing a clear breakdown of where the model succeeded and failed. For security applications, \textbf{Recall} is often the most critical metric, as missing a real attack is typically more costly than raising a false alarm. The strategic use of \textbf{Precision--Recall curves} and threshold optimization for the Probe and U2R models further underscores the prioritization of high recall for these particularly dangerous attack types.

\subsection{Attack Specific Model Performance}
\subsubsection{DoS Model Performance}
The DoS model demonstrated strong performance on the high-volume DoS attacks. The results from KDDTest are presented.

\begin{table}[h]
\centering
\caption{DoS Model Performance Metrics}
\begin{tabular}{p{3cm}p{1cm}}
\hline
\textbf{Metric} & \textbf{Value} \\
\hline
Accuracy  & 0.94 \\
Precision & 0.96 \\
Recall    & 0.90 \\
F1-Score  & 0.93 \\
\hline
\end{tabular}
\label{tab:dos_performance}
\end{table}

\subsubsection{Probe Model Performance}
The Probe model, which employed a more complex architecture and threshold optimization (threshold derived from the precision-recall curve), showed significant improvements. 

\begin{table}[h]
\centering
\caption{Probe Model Performance Metrics}
\begin{tabular}{p{3cm}p{1cm}}
\hline
\textbf{Metric} & \textbf{Value} \\
\hline
Accuracy  & 0.96 \\
Precision & 0.88 \\
Recall    & 0.93 \\
F1-Score  & 0.90 \\
\hline
\end{tabular}
\label{tab:dos_performance}
\end{table}
The model achieved a recall of 73\% with the default threshold, but by optimizing the threshold, the recall was boosted to 93\%, with a minimal change in precision. This improvement in recall translates directly to a reduction in missed attacks.

\subsubsection{R2L Model Performance}
The R2L model's performance was evaluated with a particular emphasis on its ability to detect the rare R2L attacks. The model achieved an overall accuracy of 94.53\%, but as noted, this metric can be misleading for imbalanced data. A closer look at the Classification Report reveals the true performance. For the Normal class, the model reached a precision of 97.49\% and a recall of 95.36\%. For the R2L attack class, the precision was 85.44\% and the recall was 91.75\%, resulting in an F1-score of 88.48\%. 

\begin{table}[h]
\centering
\caption{R2L Model Performance Metrics}
\begin{tabular}{p{3cm}p{1cm}}
\hline
\textbf{Metric} & \textbf{Value} \\
\hline
Accuracy  & 0.95 \\
Precision & 0.85 \\
Recall    & 0.92 \\
F1-Score  & 0.88 \\
\hline
\end{tabular}
\label{tab:dos_performance}
\end{table}
\newpage
\subsubsection{U2R Model Performance}
The U2R detection task is the most challenging due to the extreme scarcity of U2R attacks in the NSL-KDD dataset. 
The advanced U2R model achieved a ROC-AUC score of 0.9682, indicating strong discriminative ability overall. 
The confusion matrix revealed that out of 67 U2R attacks, the model successfully detected 48 (71.6\%) but missed 19 (28.4\%). 
Although the overall accuracy was high at 99.31\%, this value is largely influenced by the dominance of normal traffic and can be misleading. 

\begin{table}[h]
\centering
\caption{R2L Model Performance Metrics}
\begin{tabular}{p{3cm}p{1cm}}
\hline
\textbf{Metric} & \textbf{Value} \\
\hline
Accuracy  & 0.99 \\
Precision & 0.50 \\
Recall    & 0.72 \\
F1-Score  & 0.58 \\
\hline
\end{tabular}
\label{tab:dos_performance}
\end{table}
\subsection{Meta-Classifier for Anomaly Detection}
The specialized models trained for individual attack categories were ultimately integrated into a meta-classification framework in order to achieve robust anomaly detection. Instead of relying on a single model, this approach leverages the strengths of each specialized detector, combining their outputs as input features to a Random Forest meta-classifier. This meta-classifier then acts as the final anomaly detector, capable of capturing cross-model decision patterns while improving generalization to unseen traffic.  

\paragraph{Evaluation Strategy}  
To rigorously evaluate the generalizability of the proposed system, we tested the meta-classifier on two benchmark datasets: \texttt{KDDTest+} and \texttt{KDDTest-21}. These datasets differ in their class distributions, with \texttt{KDDTest+} being closer to the training distribution and \texttt{KDDTest-21} representing a more challenging and imbalanced scenario. Evaluating on both ensures that the meta-classifier is not only accurate under familiar conditions but also resilient when exposed to distribution shifts.  

\paragraph{Dataset Characteristics}  
The datasets contain 123 features per sample and include both majority and severely minority attack classes. Their distributions are summarized in Table~\ref{tab:test_datasets}, illustrating the strong class imbalance that makes anomaly detection challenging.  

\begin{table}[h]
\centering
\caption{Characteristics of the evaluation datasets}
\label{tab:test_datasets}
\begin{tabular}{lrr}
\hline
\textbf{Class} & \textbf{KDDTest+} & \textbf{KDDTest-21} \\
\hline
Total Samples & 22,544 & 11,850 \\
Normal        & 9,711 (43.1\%)  & 2,152 (18.2\%) \\
DoS           & 7,460 (33.1\%)  & 4,344 (36.7\%) \\
Probe         & 2,421 (10.7\%)  & 2,402 (20.3\%) \\
R2L / U2R     & 2,952 (13.1\%)  & 2,952 (24.9\%) \\
\hline
\end{tabular}
\end{table}

\paragraph{Meta-Classifier Performance.}  
The Random Forest meta-classifier demonstrated consistently strong performance across both datasets, as shown in Table~\ref{tab:meta_classifier_perf}. On \texttt{KDDTest+}, the classifier achieved near-perfect ROC-AUC (0.9995) and F1-score (0.9958), indicating its ability to detect both majority and minority attacks with high reliability. On the more challenging \texttt{KDDTest-21}, which has a higher proportion of minority classes such as R2L and U2R, the model maintained high performance, achieving an ROC-AUC of 0.9974.  

\begin{table}[h]
\centering
\caption{Meta-classifier performance on benchmark test sets}
\label{tab:meta_classifier_perf}
\begin{tabular}{lcc}
\hline
\textbf{Metric} & \textbf{KDDTest+} & \textbf{KDDTest-21} \\
\hline
Accuracy   & 0.9952 & 0.9910 \\
Precision  & 0.9933 & 0.9914 \\
Recall     & 0.9982 & 0.9976 \\
F1-Score   & 0.9958 & 0.9945 \\
ROC-AUC    & 0.9995 & 0.9974 \\
\hline
\end{tabular}
\end{table}

\begin{figure}[h!]
    \centering
    \includegraphics[width=0.6\columnwidth]{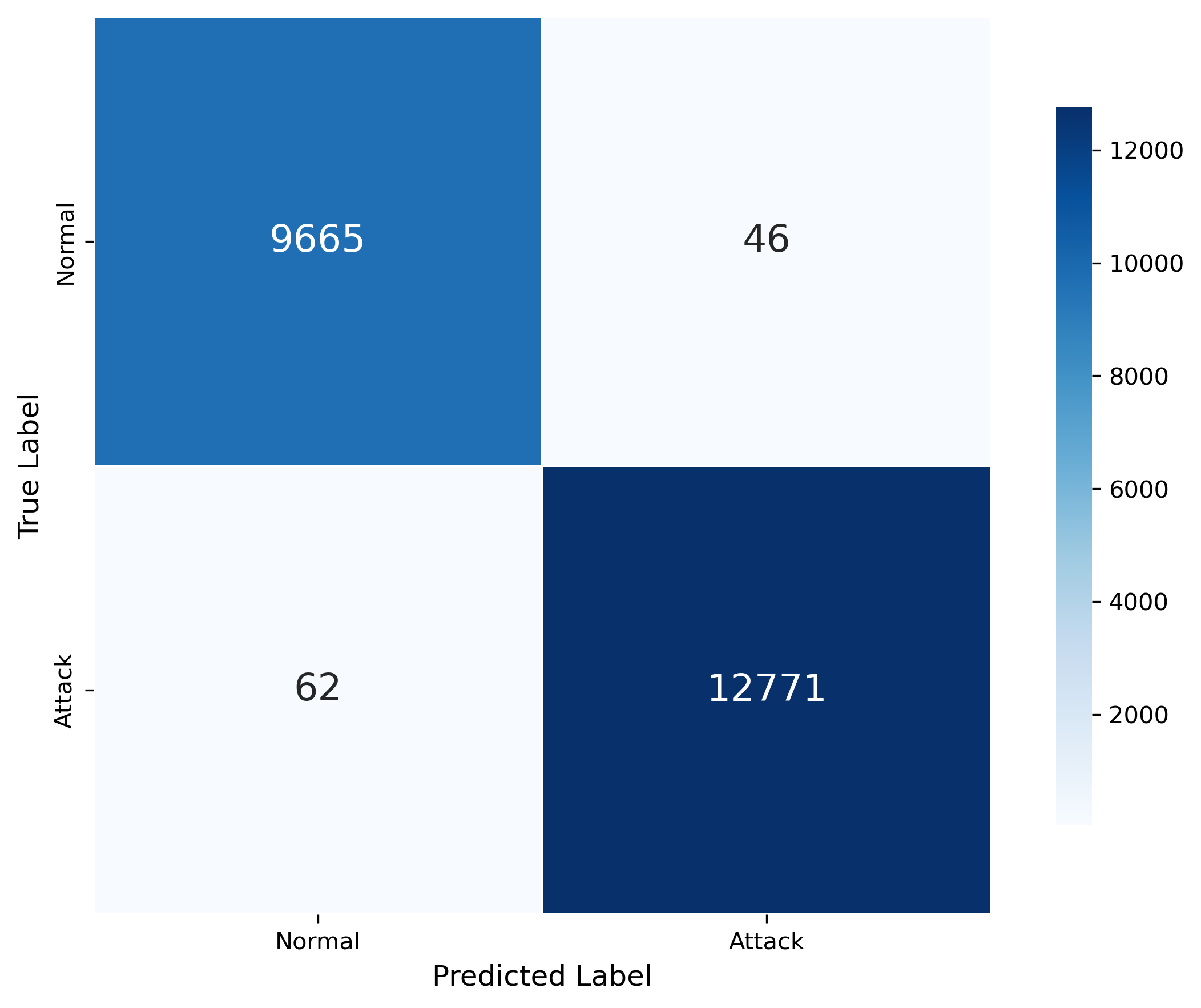}
    \caption{Confusion Matrix for KDDTest+}
    \label{fig:conf_matrix1}
    
    \includegraphics[width=0.6\columnwidth]{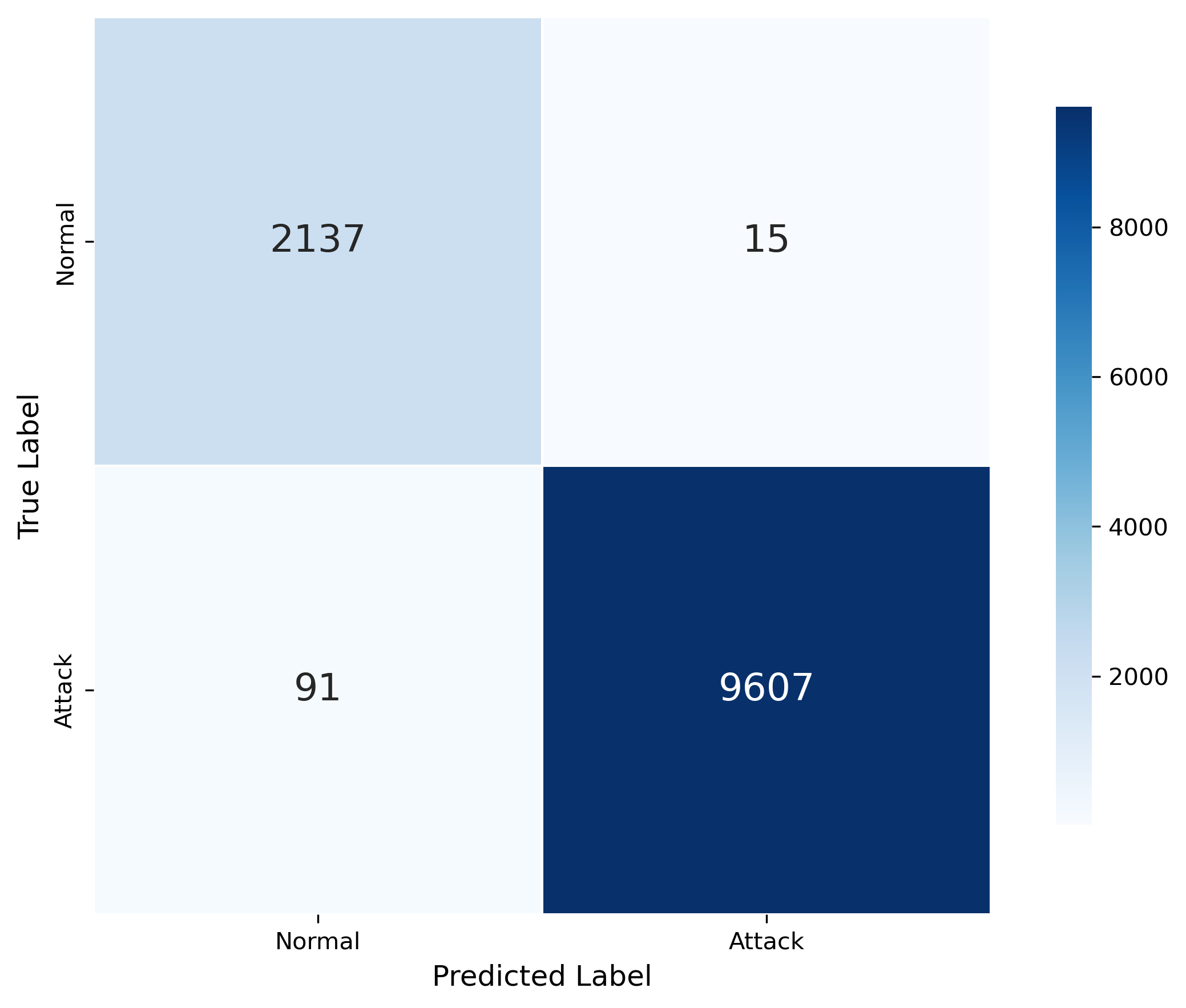}
    \caption{Confusion Matrix for KDDTest-21}
    \label{fig:conf_matrix2}
\end{figure}

\subsection{Performance Comparison of Different Methods}

To benchmark the effectiveness of the proposed hybrid framework, the meta-classifier's performance is compared against results from other recent state-of-the-art methods evaluated on the NSL-KDD dataset. All the mentioned studies have evaluated their anomaly detection performance using KDDTest+ dataset.

\begin{table}[h]
\centering
\setlength{\tabcolsep}{6pt} 
\renewcommand{\arraystretch}{1.4}
\caption{Comparative Performance on NSL-KDD}
\label{tab:comparative_performance}
\begin{tabular}{lccc}
\hline
\textbf{Study} & \textbf{Accuracy} & \textbf{Precision} & \textbf{Recall} \\
\hline
CNN-LSTM\cite{bamber_hybrid_2025} & 0.9500 & 0.9800 & 0.8900 \\
BO-GP\cite{masum_bayesian_2021} & 0.8295 & 0.7973 & 0.8135 \\
Attention-CNN-LSTM\cite{alashjaee_deep_2025} & 0.9480 & 0.9370 & 0.9250 \\
Hybrid CNN-LSTM\cite{aljanabi_effective_2024} & 0.9800 & 0.9750 & 0.9820 \\
\textbf{Ours} & \textbf{0.9952} & \textbf{0.9933} & \textbf{0.9982} \\
\hline
\end{tabular}
\end{table}

\section{DISCUSSION}
The empirical results across all four attack types strongly support the central thesis of this work: an attack-specialized framework is superior to a single, monolithic model for NIDS. By tailoring the data processing, feature selection, and model architecture to the unique characteristics of each threat, the system achieved high performance even on the most difficult tasks. The DoS model, designed for high-volume attacks, demonstrated high precision. The Probe model, with its advanced Attention and Bidirectional layers, showed a remarkable ability to detect subtle reconnaissance attempts. Most importantly, the R2L and U2R models, which faced extreme class imbalance, achieved strong recall scores by employing a combination of synthetic data generation and specialized loss functions.

This work provides a compelling comparison of various imbalance handling techniques. For DoS, SMOTE and focal loss proved effective. The Probe model's success can be directly attributed to the use of an "enhanced focal loss" and, critically, the post-training threshold optimization on the precision-recall curve. This highlights a nuanced understanding of real-world security needs, where a high Recall for a critical attack type is often more important than overall Accuracy. The R2L model’s performance validates the effectiveness of cost-sensitive learning as a viable strategy to penalize misclassifications of rare classes. The U2R model represents the culmination of these efforts, demonstrating that for extremely rare classes, a multi-faceted approach combining multiple oversampling variants (SMOTE, ADASYN, BorderlineSMOTE) and an "extreme focal loss" function is required to achieve high detection rates.

\section{CONCLUSION}
This research successfully developed and evaluated an attack-specialized hybrid framework for network intrusion detection using the NSL-KDD dataset. The methodology demonstrates that a single, standardized preprocessing pipeline is a foundational step for a reproducible system. By training dedicated, specialized models for each attack type and employing tailored techniques to handle each unique challenge, the framework achieves high detection performance across the board. The strategic use of advanced deep learning architectures, oversampling methods, and specialized loss functions, coupled with a deep understanding of domain-specific metrics like recall and precision, highlights the project's novel contribution to the field. The results strongly suggest that this specialized, modular approach is a robust and effective strategy for building next-generation intrusion detection systems capable of identifying both high-volume and stealthy, low-volume attacks.

While the results are promising, it is important to acknowledge the limitations of using a static, albeit benchmark, dataset like NSL-KDD. Real-world network traffic is dynamic and non-stationary, meaning new attack types and normal traffic patterns are constantly emerging. Furthermore, while the latency for real-time inference in similar advanced models can be low (e.g., sub-35ms for processing over 1200 records per second), achieving truly immediate responses remains a persistent engineering challenge for production NIDS. Future work should focus on validating this framework on more contemporary datasets or in a real-time environment where fast, **real-time responses** are critically needed to mitigate active threats. Additionally, while the notebooks describe the training of the specialized models, the implementation of the final meta-classifier is conceptual. Future research should focus on the design and evaluation of this final decision-making layer, which is crucial for the framework's overall utility.

\bibliographystyle{IEEEtran}
\bibliography{cyber}
\end{document}